\documentclass[preprintnumbers,prd,showpacs,floatfix,superscriptaddress,nofootinbib,twocolumn]{revtex4-1}

\pdfoutput=1
\usepackage{graphicx,epsfig}
\usepackage{amsmath}
\usepackage {amssymb}
\usepackage {longtable}
\usepackage{multirow}

\usepackage{dcolumn}
\usepackage{bm}
\usepackage{amsfonts}
\usepackage{subfigure}
\usepackage{color}

\usepackage{relsize}

\def\be{\begin{equation}}
\def\ee{\end{equation}}
\def\beq{\begin{eqnarray}}
\def\eeq{\end{eqnarray}}

\begin{document}

\title{Wormholes in generalized hybrid metric-Palatini gravity obeying the
matter null
energy condition everywhere}

\author{Jo\~{a}o Lu\'{i}s Rosa}
\email{joaoluis92@gmail.com}
\affiliation{Centro de Astrof\'isica e Gravita\c c\~ao - CENTRA,
Departamento de F\'isica,
Instituto Superior T\'{e}cnico - IST,
Universidade de Lisboa - UL,
Avenida Rovisco Pais 1, 1049-001, Portugal}

\author{Jos\'{e} P. S. Lemos}
\email{joselemos@ist.utl.pt}
\affiliation{Centro de Astrof\'isica e Gravita\c c\~ao - CENTRA,
Departamento de F\'isica,
Instituto Superior T\'{e}cnico - IST,
Universidade de Lisboa - UL,
Avenida Rovisco Pais 1, 1049-001, Portugal}

\author{Francisco S. N. Lobo}
\email{fslobo@fc.ul.pt}
\affiliation{Instituto de Astrof\'{\i}sica e Ci\^{e}ncias do
Espa\c{c}o, Faculdade de Ci\^encias, Universidade de Lisboa - UL,
Edif\'{\i}cio C8, Campo Grande, P-1749-016 Lisbon, Portugal}


\begin{abstract} 
Wormhole solutions in a generalized hybrid metric-Palatini matter theory, given by a gravitational Lagrangian $f\left(R,\cal{R}\right)$, where $R$ is the metric Ricci scalar, and $\mathcal{R}$ is a Palatini scalar curvature defined in terms of an independent connection, and a matter Lagrangian, are found. The solutions are worked in the scalar-tensor representation of the theory, where the Palatini field is traded for two scalars, $\varphi$ and $\psi$, and the gravitational term $R$ is maintained. The main interest in the solutions found is that the matter field obeys the null energy condition (NEC) everywhere, including the throat and up to infinity, so that there is no need for exotic matter. The wormhole geometry with its flaring out at the throat is supported by the higher-order curvature terms, or equivalently, by the two fundamental scalar fields, which either way can be interpreted as a gravitational fluid. Thus, in this theory, in building a wormhole, it is possible to exchange the exoticity of  matter by the exoticity of the gravitational sector. The specific wormhole displayed, built to obey the matter NEC from the throat to infinity, has three regions, namely, an interior region containing the throat, a thin shell of matter, and a vacuum Schwarzschild anti-de Sitter (AdS) exterior. For hybrid metric-Palatini matter theories this wormhole solution is the first where the NEC for the matter is verified for the entire spacetime keeping the solution under asymptotic control.  The existence of this type of solutions is in line with the idea that traversable wormholes bore by additional fundamental gravitational fields, here disguised as scalar fields, can be found without exotic matter. Concomitantly, the somewhat concocted architecture needed to assemble a complete wormhole solution for the whole spacetime may imply that in this class of theories such solutions are scarce.
\end{abstract}


\maketitle

\section{Introduction}\label{Introduction}

Within general relativity, wormholes were found as exact solutions
connecting two different asymptotically flat regions of spacetime
\cite{Morris:1988cz,Visser:1995cc} as well as two different
asympotically de Sitter (dS) or anti-de Sitter (AdS) regions
\cite{lemoslobooliveira}.  The fundamental ingredient in wormhole
physics is the existence of a throat satisfying a flaring-out
condition. In general relativity this geometric condition entails the
violation of the null energy condition (NEC).
This NEC
states that $T_{ab}\,k^a k^b \geq 0$, where $T_{ab}$ is the matter
stress-energy tensor and $k^a$ is any null vector. Matter that
violates the NEC is denoted as exotic
matter. Wormhole solutions have been also found in other theories,
see, e.g., \cite{Camera:1995,Nandi:1998,Bronnikov:2002rn,Camera:2003,
Lobo:2007zb,Garattini:2007ff,Lobo:2008,Garattini:2009,Lobo:2010sb,
Garattini:2011fs} and \cite{Lobo:2017oab} for reviews.  In these
works the NEC for
the matter is also violated.  However due to its
nature, it is important and useful to minimize its usage.

In fact, in the context of modified theories of gravity, it has been
shown in principle that normal matter may thread the wormhole throat,
and it is the higher-order curvature terms, which may be interpreted
as a gravitational fluid, that support these nonstandard wormhole
geometries.  Indeed, in \cite{Lobo:2009ip} it was shown explicitly
that in $f(R)$ theories wormhole throats can be theoretically
constructed without the presence of exotic matter, in
\cite{Garcia:2010xb,MontelongoGarcia:2010xd} nonminimal couplings were
used to build such wormholes, and in \cite{Harko:2013yb} generic
modified gravities were used also with that aim in mind, i.e., the
wormhole throats are sustained by the fundamental fields presented in
the modified gravity alone.  This type of solutions were also found in
Einstein-Gauss-Bonnet theory
\cite{Bhawal:1992,Dotti:2007,Mehdizadeh:2015jra}, and are mentioned in
brane world scenarios \cite{Lobo:2007}, in Brans-Dicke theories
\cite{Anchordoqui:1997}, and in a hybrid metric-Palatini gravitational
theory \cite{Capozziello:2012hr}.  It is our aim to find wormhole
solutions whose matter obeys the NEC not only at the
throat but everywhere in a generalized hybrid metric-Palatini gravity,
with action $f(R,{\cal R})$.

The action $f(R,{\cal R})$ is well motivated.  Indeed, a promising
approach to modified gravity consists in having a hybrid
metric-Palatini gravitational theory \cite{Harko:2011nh}, which
consists in adding to the Einstein-Hilbert action $R$, a new term
$f(\cal{R})$, where $\cal{R}$ is a curvature scalar defined in terms
of an independent connection, and $f$ is some function of $\cal{R}$.
In this approach, the metric and affine connection are regarded as
independent degrees of freedom, and contrary to general relativity
where the metric-affine, or Palatini, formalism coincides with the
purely metric formalism, in an $R+f({\cal R})$ theory the two
formalisms lead to different results \cite{Olmo:2011uz}.  In the
$R+f({\cal R})$ theory one retains through $R$ the positive results of
general relativity, with further gravitational degrees of freedom
being represented in the metric-affine $f({\cal R})$ component.  One
can further express the $R+ f({\cal R})$ theory in a dynamically
equivalent scalar-tensor representation which simplifies the
analysis. In this representation, besides wormhole solutions
\cite{Capozziello:2012hr}, solar system tests and cosmological
solutions have been analyzed and found
\cite{Capozziello:2012ny,Capozziello:2012qt,Capozziello:2013yha,
Capozziello:2013uya,Capozziello:2015lza}, see also \cite{HarkoLobo}
for a review.  A natural generalization to $R+f({\cal R})$ is to
consider an $f(R,{\cal R})$ action, i.e., the gravitational action is
taken to depend on a general function of both the metric and Palatini
curvature scalars \cite{Tamanini:2013ltp}.  One can also find the
scalar-tensor representation of this generalized hybrid
metric-Palatini gravity, which now has two scalar fields.  Exact
solutions were constructed representing an FLRW
(Friedmann-Lema\^itre-Robsertson-Walker) universe in a generalized
hybrid metric-Palatini theory \cite{Rosa:2017jld}.  Among other
relevant results, it was shown that it is possible to obtain
exponentially expanding solutions for flat universes even when the
cosmology is not purely vacuum. In addition, the junction conditions
in this theory have been worked out \cite{rosalemos1}.

In this work, our aim is to find static and spherically symmetric
wormholes solutions in the generalized $f(R,{\cal R})$ hybrid
metric-Palatini matter theory in which the matter satisfies
the NEC
everywhere, from the throat to infinity, so there is no need
for exotic matter. This fills a gap in the literature as most
of the work that has been done in this area has been aimed at finding 
solutions where the NEC is satisfied solely at
the wormhole throat paying no attention to the other regions. As
far as we are aware our work is the first where the
NEC
is
verified for the entire spacetime.

This paper is organized as follows. In
Sec.~\ref{secII}, we consider the action and write out the
gravitational field equations, both in the curvature and in the
equivalent scalar-tensor representation. In Sec.~\ref{secIII}, we
present the equations of motion for static and spherically symmetric
wormholes solutions. In Sec.~\ref{secIV}, we impose specific choices
for the metric redshift and shape functions and
for the potential governing the scalar fields, and find
solutions where the matter threading the wormhole 
satisfies the NEC in the 
whole spacetime. In Sec.~\ref{conclusion}, we conclude.

\section{Generalized hybrid metric-Palatini gravity with matter and its scalar
representation}\label{secII}

\subsection{Action and field equations}

Consider the general hybrid action $S$ given by
\be\label{genac}
S=\frac{1}{2\kappa^2}\int_\Omega\sqrt{-g}f\left(R,\cal{R}\right)d^4x+
\int_\Omega\sqrt{-g}\;{\cal L}_m d^4x,
\ee
where $\kappa^2 \equiv 8\pi G$,
$G$ is the gravitational constant, the speed of light is
put to one, $\Omega$ is the spacetime volume, 
$g$ is the determinant of the spacetime
metric $g_{ab}$, latin indices $a,b$ run from 0 to 3, 
$R$ is the metric Ricci scalar,
$\mathcal{R}\equiv\mathcal{R}^{ab}g_{ab}$ is the Palatini Ricci scalar,
where the Palatini Ricci tensor
is defined in terms of an independent connection $\hat\Gamma^c_{ab}$ as, 
$
\mathcal{R}_{ab}=\partial_c
\hat\Gamma^c_{ab}-\partial_b\hat\Gamma^c_{ac}+\hat\Gamma^c_{cd}
\hat\Gamma^d_{ab}-\hat\Gamma^c_{ad}\hat\Gamma^d_{cb}
$, $\partial_a$ denotes partial derivative,
$f\left(R,\cal{R}\right)$ is a well behaved
function of $R$ and $\cal{R}$, and
${\cal L}_m$ is the matter Lagrangian density
considered minimally coupled to the metric $g_{ab}$.

Variation of the action  (\ref{genac})
with respect to the metric $g_{ab}$ yields the following equation of motion
\beq
\frac{\partial f}{\partial R}R_{ab}+\frac{\partial f}{\partial \mathcal{R}}\mathcal{R}_{ab}-\frac{1}{2}g_{ab}f\left(R,\cal{R}\right)
   \nonumber \\
-\left(\nabla_a\nabla_b-g_{ab}\Box\right)\frac{\partial f}{\partial R}=\kappa^2 T_{ab},
\eeq
where $\nabla_a$ is the covariant derivative and $\Box=\nabla^a\nabla_a$ the d'Alembertian,
both with respect to $g_{ab}$, and
$T_{ab}$ is the stress-energy tensor defined in the usual manner as
\begin{equation}
T_{ab}=-\frac{2}{\sqrt{-g}}\frac{\delta(\sqrt{-g}\,{\cal
L}_m)}{\delta(g^{ab})} ~.
 \label{defSET}
\end{equation}

Varying the action (\ref{genac}) with respect to the independent
connection $\hat\Gamma^c_{ab}$
provides the following relationship
\be
\hat\nabla_c\left(\sqrt{-g}\frac{\partial f}{\partial \cal{R}}g^{ab}\right)=0 \,,
\label{eqvar1}
\ee
where $\hat\nabla_a$ is the covariant derivative with respect to the connection
$\hat\Gamma^c_{ab}$.
Now, recalling that $\sqrt{-g}$ is a scalar density of weight
1, we have that $\hat\nabla_c \sqrt{-g}=0$ and so Eq.~(\ref{eqvar1}) simplifies to
$\hat\nabla_c\left(\frac{\partial f}{\partial \cal{R}}g^{ab}\right)=0$. This means that $h_{ab}$
defined as
\be
h_{ab}=g_{ab} \frac{\partial f}{\partial \cal{R}} \,,
\label{hab}
\ee
is a metric to the connection $\hat\Gamma^a_{bc}$
which then can be written as the following Levi-Civita connection 
\be
\hat\Gamma^a_{bc}=\frac{1}{2}h^{ad}\left(\partial_b h_{dc}+\partial_c h_{bd}-\partial_d h_{bc}\right)\,.
\ee
Note also from Eq.~(\ref{hab}) that $h_{ab}$ is conformally related to $g_{ab}$ through the
conformal factor ${\partial f}/{\partial \cal{R}}$.

\subsection{Scalar-tensor representation of the generalized hybrid metric-Palatini gravity with matter}\label{str}

In order to have a better grasp on the sytem of equations, 
it is useful to express the action \eqref{genac} in a scalar-tensor representation. This can be achieved by considering an action with two auxiliary fields, $\alpha$ and $\beta$, respectively, in the following form
\beq
S=\frac{1}{2\kappa^2}\int_\Omega \sqrt{-g}\Big[f\left(\alpha,\beta\right)+\frac{\partial f}{\partial \alpha}\left(R-\alpha\right)
    \nonumber  \\
+\frac{\partial f}{\partial\beta}\left(\cal{R}-\beta\right)\Big]d^4x+\int_\Omega\sqrt{-g}\;{\cal L}_m d^4x.
\label{gensca}
\eeq
Using $\alpha=R$ and $\beta=\mathcal{R}$ we recover action \eqref{genac}. Therefore, we can define two scalar fields as $\varphi=\partial f(\alpha,\beta)/\partial\alpha$ and $\psi=-\partial f(\alpha,\beta)/\partial\beta$, where the negative sign is imposed here for convention. The equivalent action is of the form
\beq
S=\frac{1}{2\kappa^2}\int_\Omega \sqrt{-g}\left[\varphi R-\psi\mathcal{R}-V\left(\varphi,\psi\right)\right]d^4x
	\nonumber \\
+\int_\Omega\sqrt{-g}\;{\cal L}_m d^4x,
  \label{action3}
\eeq
where we defined the potential $V\left(\varphi,\psi\right)$ as
\be\label{potential}
V\left(\varphi,\psi\right)=-f\left(\alpha,\beta\right)+\varphi\alpha-\psi\beta,
\ee
and here $\alpha$ and $\beta$ should be seen as functions
of $\varphi$ and $\psi$.
Taking into account that $h_{ab}$ defined in
Eq.~(\ref{hab}) is conformally related to $g_{ab}$
and that it
can now be written as
$h_{ab}=-\psi\, g_{ab}$, we have that the Ricci scalars
$\mathcal{R}$ and $R$ are related through
\be\label{confrt}
\mathcal{R}=R+\frac{3}{\psi^2}\partial^a \psi\partial_a \psi-
\frac{3}{\psi}\Box\psi\,.
\ee
So, we can replace $\cal{R}$ in the action (\ref{action3}), 
to obtain
\beq\label{genacts2}
S=\frac{1}{2\kappa^2}\int_\Omega \sqrt{-g}\Big[\left(\varphi-\psi\right) R
-\frac{3}{2\psi}\partial^a\psi\partial_a\psi
     \nonumber  \\
-V\left(\varphi,\psi\right)\big]d^4x+\int_\Omega\sqrt{-g}\;{\cal L}_m d^4x.
\eeq

Varying the action \eqref{genacts2} with respect to the metric $g_{ab}$
provides the following gravitational equation 
\beq
&&\left(\varphi-\psi\right) G_{ab}=\kappa^2T_{ab}
+\nabla_a\nabla_b
\varphi-\nabla_a\nabla_b\psi
+\frac{3}{2\psi}\partial_a\psi\partial_b\psi
	\nonumber  \\
&&-\left(\Box\varphi-\Box\psi+\frac{1}{2}V+\frac{3}{4\psi}
\partial^c\psi\partial_c\psi\right)g_{ab}\,.
\label{genein2}
\eeq

Varying the action with respect to the field $\varphi$ and to the 
field $\psi$ yields after rearrangements, see \cite{Rosa:2017jld},
\be\label{genkgi}
\Box\varphi+\frac{1}{3}\left(2V-\psi V_\psi-\varphi V_\varphi\right)
=\frac{\kappa^2T}{3}\,,
\ee
and
\be\label{genkg}
\Box\psi-\frac{1}{2\psi}\partial^a\psi\partial_a\psi-\frac{\psi}{3}
\left(V_\varphi+V_\psi\right)=0\,,
\ee
respectively.

In the next section, we will use the equations of motion
(\ref{genein2}), (\ref{genkgi}) and (\ref{genkg}) to find static and
spherically symmetric wormhole solutions.  We will do two things. We
will impose the flaring out condition that has to be obeyed at a
wormhole throat, and we impose that the NEC
on the matter stress-energy tensor are obeyed.

\section{Wormhole ansatz and equations}\label{secIII}

\subsection{Wormhole ansatz and equations}

We impose that the 
wormhole solutions are described
by a static and spherically symmetric metric 
which in the usual spherical $(t,r,\theta,\phi)$
coordinates has components 
$g_{ab}={\rm diag}\,
(g_{tt},g_{rr},g_{\theta\theta},g_{\phi\phi})$, and
whose corresponding line element has then the
form \cite{Morris:1988cz}
\be
\label{worm}
ds^2=-e^{\zeta\left(r\right)}dt^2+
\left[1-\frac{b\left(r\right)}{r}\right]^{-1}dr^2+r^2
\left( d\theta^2+\sin^2\theta d\phi^2\right)\,,
\ee
where  $\zeta\left(r\right)$ is the redshift function and
$b\left(r\right)$ is the shape function. 
The shape function $b(r)$
should obey the boundary condition
$
b\left(r_0\right)=r_0$,
where $r_0$ is
the radius of the wormhole throat.

Since the geometry is static and spherically symmetric,
we assume that the scalar fields 
are functions of the radial coordinate alone,
\be
\label{varphir}
\varphi=\varphi(r)\,,
\ee
\be
\label{psir}
\psi=\psi(r)\,.
\ee

We further assume
that matter is described by an anisotropic stress-energy tensor of the form
\be\label{tab}
{T_a}^b={\rm diag }\left(-\rho,p_r,p_t,p_t\right)\,,
\ee
where $\rho=\rho(r)$ is the energy density, $p_r=p_r(r)$
is the radial pressure, and 
$p_t=p_t(r)$
is the transverse pressure.
We can now use the
ans\"atze given in
Eqs.~(\ref{worm})-(\ref{tab})
to derive from Eqs.~(\ref{genein2})-(\ref{genkg})
the system of equations appropriate to this problem.

The three independent components of Eq.~\eqref{genein2} are given by
\beq\label{genfieldr}
&&\left(\varphi-\psi\right)\frac{b'}{r^2}
-\frac{\left(\varphi-\psi\right)'}{2r}\left(1-b'\right)-\frac{V}{2}
	\\
&-&\left(1-\frac{b}{r}\right)\left[\left(\varphi-\psi\right)''+\frac{3\left(\varphi-\psi\right)'}{2r}+\frac{3\psi'^2}{4\psi}\right]=\kappa^2\rho\,,\nonumber 
\eeq
\beq\label{genfieldpr}
&&\left(\varphi-\psi\right)\left[\frac{\zeta'}{r}\left(1-\frac{b}{r}\right)-\frac{b}{r^3}\right]
+\left(1-\frac{b}{r}\right)\times
	 \\
&\times&
\left[-\frac{3\psi'^2}{4\psi}+\frac{2\left(\varphi-\psi\right)'}{r}+\frac{\zeta'\left(\varphi-\psi\right)'}{2}\right]+\frac{V}{2}=\kappa^2p_r,\nonumber
\eeq
\beq\label{genfieldpt}
&&\left(\varphi-\psi\right)\Bigg[\left(1-\frac{b}{r}\right)\left(\frac{\zeta''}{2}+\frac{\zeta'^2}{4}+\frac{\zeta'}{2r}\right)
	\\
&&+\frac{b-rb'}{2r^3}\left(1+\frac{\zeta'r}{2}\right)\Bigg]+\frac{\phi'}{2r}\left(1-b'\right)+\frac{V}{2}+\left(1-\frac{b}{r}\right)\times
	 \nonumber \\
&&\times \left[\left(\varphi-\psi\right)''+\frac{\zeta'\left(\varphi-\psi\right)'}{2}+\frac{3\psi'^2}{4\psi}+\frac{\left(\varphi-\psi\right)'}{2r}\right]=\kappa^2p_t \,.\nonumber 
\eeq

The scalar field equation
for $\varphi(r)$,
Eq.~\eqref{genkgi},
after rearrangements,  can be written as
\beq
\left(1-\frac{b}{r}\right)\left(\zeta''+\frac{\zeta'^2}{2}+\frac{2\zeta'}{r}\right)
 -\frac{b+rb'}{r^3}
\nonumber \\
+\frac{b-rb'}{r^3}\left(1+\frac{r\zeta'}{2}\right)+V_\varphi=0 \,.
 \label{genkgtrace}
\eeq
The scalar field equation
for $\psi(r)$,
Eq.~\eqref{genkg},
can be written as
\beq\label{genwormkg1}
\left(1-\frac{b}{r}\right)\left(\psi''+\frac{\zeta'\psi'}{2}+\frac{3\psi'}{2r}-\frac{\psi'^2}{2\psi}\right)
	\nonumber \\
+\frac{\psi'}{2r}\left(1-b'\right)-\frac{\psi}{3}\left(V_\varphi+V_\psi\right)=0\,.
\eeq
To derive Eq.~(\ref{genkgtrace}) we do the following.
Equation \eqref{genkgi} gives directly an equation for $\varphi''$, i.e., 
$
\left(1-\frac{b}{r}\right)\left(\varphi''+\frac{\zeta'\varphi'}{2}+
\frac{3\varphi'}{2r}\right)+\frac{\varphi'}{2r}\left(1-b'\right)
+\frac{1}{3}\left(2V-\psi V_\psi-\varphi V_\varphi\right)=\frac{\kappa^2T}{3}
$.
On the other hand, the trace of the stress-energy tensor $T$
is $T=-\rho+p_r+2p_t$. From
Eqs.~\eqref{genfieldr}, \eqref{genfieldpr}, and \eqref{genfieldpt} we get
$
\kappa^2T = \left(\varphi-\psi\right)\left[\left(1-\frac{b}{r}
\right)\left(\zeta''+\frac{\zeta'^2}{2}+\frac{2\zeta'}{r}\right)
+\frac{b-rb'}{r^3}\left(1+\frac{r\zeta'}{2}\right)\right.-
\newline\break
\left.\frac{b+rb'}{r^3}\right]
+3\left(1-\frac{b}{r}\right)\left[\left(\varphi-\psi\right)''+
\frac{3\left(\varphi-\psi\right)'}{2r}+\frac{\psi'^2}{2\psi}
+\frac{\zeta'\left(\varphi-\psi\right)'}{2}\right]+
\frac{3\left(\varphi-\psi\right)'}{2r}\left(1-b'\right)+2V$.
Subtracting Eq.~\eqref{genwormkg1} to
the $\varphi''$ equation
and replacing in the resulting
equation the value of $T$ just obtained,
the terms dependent on $\psi$ and $\varphi$ cancel out,
yielding Eq.~(\ref{genkgtrace}).

We have now a system of equations to solve. There are five independent
equations, given by the three field equations
\eqref{genfieldr}-\eqref{genfieldpt}, and the two scalar field
equations \eqref{genkgtrace} and \eqref{genwormkg1}. On the other
hand, there are eight variables, $\zeta(r)$, $b(r)$, $\varphi(r)$,
$\psi(r)$,
$V(r)$, $\rho(r)$, $p_r(r)$, and $p_t(r)$.  Therefore, we
can give three functions to the system to solve it.

\subsection{Strategy for wormhole solutions obeying the matter null energy condition}

\subsubsection{Main aim: Wormhole matter fields do not violate the null energy condition anywhere}

The main aim in our wormhole construction
is that throughout the wormhole solution
the matter must obey the NEC. This needs
some explanation. In pure general
relativity, it is known that at a wormhole troat,
the NEC for the matter
fields is violated in order to
allow for the wormhole flare out. In the
scalar representation of 
generalized hybrid metric-Palatini gravity
we are studying,
the gravitational field
is complemented by the two other fundamental
gravitational fields,
namely, the scalar fields
$\varphi$ and $\psi$, whereas 
the matter fields in this theory are
still encoded in the stress-energy tensor $T_{ab}$.
Thus, to allow
for a mandatory flare out (see, e.g. \cite{Visser:1995cc}),
the matter $T_{ab}$ can obey the NEC,
as long as some appropriate combination of $T_{ab}$ with
the other fundamental fields
does not obey
the NEC. This is our main
objective: to find wormhole solutions in a theory with extra
fundamental gravitational fields, represented here
by the two scalar fields,
in which the matter obeys the NEC.

To build such a wormhole might be a difficult task as the
NEC can be obeyed at a certain radius and then be violated at
some other radius.  Our strategy is then the following.  The most
critical radius is the wormhole throat radius.  So we impose that the
NEC for the matter fields is obeyed at the throat
and its vicinity. We then look for a solution in the vicinity of the
throat. When the solution starts to violate the NEC
above a certain $r$ away from the throat, we cut the solution
there. We then join it to a
vacuum solution for larger radii
and build a thin shell solution at a junction
radius up to which
the matter fields obey the NEC.

To be specific, we start by finding wormhole solutions
for which the matter threading the
throat is normal matter, i.e., the matter stress-energy tensor
$T_{ab}k^ak^b$, for two null
vectors $k^a$ and $k^b$, must obey the NEC, i.e., 
$
T_{ab}k^ak^b>0
$.
This NEC is to be obeyed
at the throat and its vicinity.
Noting that ${T_a}^b$ is of the form given in Eq.~(\ref{tab}),
taking into account Eqs.~(\ref{genfieldr})-(\ref{genfieldpt}),
and choosing 
in the frame of Eq.~(\ref{worm})
a null vector $k^a$ of the form
$k^a=\left(1,1,0,0\right)$,
one obtains from $T_{ab}k^ak^b >0$ that
\begin{equation}
\rho+p_r>0\,.
\label{rhopr}
\end{equation}
For a null vector $k^a$ of the form $k^a=\left(1,0,1,0\right)$, we find that 
$T_{ab}k^ak^b >0$ means
\begin{equation}
\rho+p_t>0\,.
\label{rhopt}
\end{equation}
It might happen that the weak energy condition (WEC) for
the matter alone is also be obeyed
at the wormhole throat, although in
our wormhole construction we do not impose it.
To verify whether the WEC
is verified or not, we choose a timelike vector
$u^a$ of the form $u^a=\left(1,0,0,0\right)$,
and see if $T_{ab}u^au^b>0$. 
For a perfect fluid 
$T_{ab}u^au^b>0$ becomes
\begin{equation}
\rho>0\,.
\label{rho}
\end{equation}
Thus the matter fields that build the wormhole
we are constructing
must obey the NEC
Eqs.~(\ref{rhopr})-(\ref{rhopt}), and
might, but not necessarily, obey the 
WEC Eq.~(\ref{rho}).

\subsubsection{Flaring out condition at a wormhole throat}

The shape function $b(r)$
obeys two boundary conditions. First,
\begin{equation}
b\left(r_0\right)=r_0,
\label{nbr0}
\end{equation}
where $r_0$ is
the radius of the wormhole throat.  Second,
the fundamental condition in
wormhole physics is that the throat flares out which is translated by
the following condition at the throat, $(b-b'r)/b^2>0$. This imposes
that at the throat we have 
\cite{Morris:1988cz,Visser:1995cc}
\begin{equation}
b'\left(r_0\right)<1.
\label{flao}
\end{equation}

The flaring out
condition, Eq.~(\ref{flao}),
is a geometric condition, indeed
it is 
a condition on the Einstein tensor $G_{ab}$.
In general relativity, since the field equation
is $G_{ab}=\kappa^2T_{ab}$, 
it is also a  condition on the
matter stress-energy tensor $T_{ab}$
and it directly implies that $T_{ab}$
violates the NEC.
However, in this alternative theory,
i.e., the scalar field representation of the
generalized hybrid  metric-Palatini
theory, besides the metric gravitational field,
there are two extra gravitational
fundamental fields $\varphi$ and $\psi$.
So, the flaring out
condition on $G_{ab}$,
in the hybrid generalization, translates into a condition
on a combination of the matter stress-energy
$T_{ab}$ with an appropriate tensor built out of
$\varphi$, $\psi$, and geometric tensorial quantities.
Let us see this in detail. 
Note that Eq.~\eqref{genein2} can be written in the form
$
\left(\varphi-\psi\right)\left(G_{ab}+H_{ab}\right)=\kappa^2T_{ab},
$
where the tensor $H_{ab}$ is  given by
$H_{ab}=\frac{1}{\varphi-\psi}[(\Box\varphi-
\Box\psi+\frac{1}{2}V+\frac{3}{4\psi}\partial^c
\psi\partial_c\psi)g_{ab}
-\frac{3}{2\psi}\partial_a\psi\partial_b\psi-\nabla_a
\nabla_b\varphi+\nabla_a\nabla_b\psi]\,.
$
We can define for clarity an effective stress-energy
tensor $T_{ab}^{\rm eff}$
as
$T_{ab}^{\rm eff}=\frac{T_{ab}}{\varphi-\psi}-\frac{H_{ab}}{\kappa^2}
$.
Our aim, defined from the start, is 
that the NEC on the matter $T_{ab}$
is obeyed everywhere, including the
wormhole throat. But, since for
wormhole construction one needs to flare out
at the throat
then the NEC on $T_{ab}^{\rm eff}$
has to be violated there.
This means that the
contraction of the effective stress-energy tensor with two null
vectors, $k^a$ and $k^b$ say, must be negative, i.e., 
$T_{ab}^{\rm eff} \,k^ak^b<0$, at the throat. 
From
the definition of $T_{ab}^{\rm eff}$ above,
this condition can be converted into
$
T_{ab}k^ak^b-g H_{ab}\,k^ak^b<0
$,
where $g(\varphi,\psi)\equiv\frac{\varphi-\psi}{\kappa^2}$.
So it gives $T_{ab}k^ak^b<g H_{ab}\,k^ak^b$.
We assume that $g>0$ and write thus this condition as
$
T_{ab}k^ak^b < g H_{ab}k^ak^b
$,
at the throat, i.e., this 
condition has to be obeyed at the
throat.
Noting that $T_{ab}$ is of the form given in Eq.~(\ref{tab}),
taking into account Eqs.~(\ref{genfieldr})-(\ref{genfieldpt}),
and choosing 
in the frame of Eq.~(\ref{worm})
a null vector $k^a$ of the form
$k^a=\left(1,1,0,0\right)$, we find that 
$T_{ab}k^ak^b < g H_{ab}k^ak^b$
at $r_0$, the throat, is then
$
\rho+p_r <
\frac{1}{4 \kappa ^2 r^2 \psi}
\{
-\frac{2 r^2 V \psi  \left[e^{\zeta } (r-b)-r\right]}{(r-b)}
+2 \psi  
\left[ (\varphi ' -\psi ') [e^{\zeta} 
(r (b'-4)+3 b)+r (r \zeta '+4)] 
\right. \break\left.-2 e^{\zeta } r (r-b)
   (\varphi ''-\psi '')\right]
-3 r (\psi ')^2 [e^{\zeta}(r-b)+r]
\}
$.
For a null vector $k^a$ of the form $k^a=\left(1,0,1,0\right)$, we find that 
$T_{ab}k^ak^b < g H_{ab}k^ak^b$
is then
$\rho+p_t <
\frac{1}{4 \kappa ^2 r^2 \psi}
\{
[2rV\psi+3(r-b)\psi'^2]
r(r^2-e^\zeta)
-2 \psi  
\left[
(\varphi ' -\psi ')
[r^2(b+r(b'-2+(b-r)\zeta'))
\right.\break\left.
-e^\zeta\left(3b+\left(b'-4\right)r\right)]+2r
(e^\zeta-r^2)(r-b)(\varphi''-\psi'')
\right]
\}
$.
The violation of the NEC
implies the
violation of the WEC
at the wormhole throat
\cite{Visser:1995cc}. We choose a timelike vector
$u^a$ of the form $u^a=\left(1,0,0,0\right)$, and we find that
$T_{ab}u^au^b < g H_{ab}u^au^b$ becomes
$ \rho <
\frac{1}{4 \kappa ^2 r^2 \psi}
\{
2r^2V\psi+3r\left(r-b\right)\psi'^2+
2\psi\left[\left(\varphi'-\psi'\right)\left(4\left(4-b'\right)-3b\right)+2r
\left(r-b\right)\left(\varphi''-\psi''\right)\right]
\}
$. 
These inequalities
at the wormhole throat, coming out of 
$T_{ab}^{\rm eff}$ to enable flaring
out, and the NEC and WEC
on $T_{ab}$,
Eqs.~(\ref{rhopr})-(\ref{rho}), 
are not incompatible at all.

So, it is possible
in principle to find wormhole solutions
in the generalized hybrid theory obeying the
matter NEC  given in
Eqs.~(\ref{rhopr})-(\ref{rhopt}).

\subsubsection{General remarks on the behavior of the fundamental fields}

The fundamental gravitational fields are the metric fields, namely,
the redshift function $\zeta\left(r\right)$, the shape function
$b\left(r\right)$, and the scalar fields $\varphi\left(r\right)$ and
$\psi\left(r\right)$ of which the potential $V$ is formed.

The redshift function $\zeta$ must be finite everywhere so that the
term $g_{tt}=e^{\zeta}$ in the metric \eqref{worm} is always finite
and does not change sign avoiding thus the presence of horizons.  The
shape function $b$ essentially should be well-behaved throughout $r$
and obey the flare out condition at the throat $r_0$ as we discussed
above, see \eqref{flao}.  The scalar fields $\varphi$ and $\psi$
should be well-behaved throughout $r$ and the potential $V$, dependent
directly on them which in turn depend on $r$, should be chosen
appropriately.

\section{Wormhole solutions in the generalized hybrid theory with matter obeying
the null energy condition everywhere}\label{secIV}

\subsection{Patch by patch building the wormhole solution}

\subsubsection{Wormhole solutions in the neighborhood of the throat}

We have the freedom to choose three functions out of 
the 
eight variables, $\zeta(r)$, $b(r)$, $\varphi(r)$,
$\psi(r)$,
$V(r)$, $\rho(r)$, $p_r(r)$, and $p_t(r)$, since
there are only
 five independent
equations
\eqref{genfieldr}-\eqref{genkgtrace}. 
The three functions that we choose are 
$\zeta(r)$, $b(r)$, and $V(\varphi,\psi)$.

A quite general class of redshift $\zeta\left(r\right)$
and shape $b\left(r\right)$
metric functions, see  Eq.~(\ref{worm}),
that verify the traversability
conditions for a wormhole is
\beq
\zeta\left(r\right)&=&\zeta_0\left(\frac{r_0}{r}
\right)^\alpha\exp\left(-\gamma\frac{r-r_0}{r_0}\right),
\label{redshift1}
   \\
b\left(r\right)&=& r_0\left(\frac{r_0}{r}
\right)^\beta\exp\left(-\delta\frac{r-r_0}{r_0}\right)\,,
\label{shape11}
\eeq
where $\zeta_0$ is a constant with no units, $r_0$ is
the wormhole throat radius, and
$\alpha$, $\gamma$, $\beta$, and $\delta$, are free exponents. 
We also assume that $\gamma>0$ so that
the redshift function $\zeta(r)$ given in Eq.~(\ref{redshift1})
is finite and nonzero for every value of $r$ between $r_0$ and infinity, and 
$\zeta_0$ is the value of $\zeta$ at $r_0$.
We further assume that $\delta\geq0$,
so that the shape
function $b(r)$
given in Eq.~(\ref{shape11}) yields
$b\left(r=r_0\right)=r_0$ 
and $b'\left(r_0\right)<1$, 
as it should for a wormhole solution,
see Eqs.~(\ref{nbr0}) and (\ref{flao}).
We consider further a power-law potential of the form 
\be
V\left(\varphi,\psi\right)=V_0\left(\varphi-\psi\right)^\eta\,,
\label{powerpotential}
\ee
where $V_0$ is a constant and $\eta$ some exponent.
With these choices of Eqs.~(\ref{redshift1})-(\ref{powerpotential})
and for various
combinations of the parameters $\alpha, \beta, \gamma, \delta$. and
$\eta$ one can then find specific equations for $\varphi$
and $\psi$ from Eqs.~\eqref{genkgtrace} and~\eqref{genwormkg1}
and try analytical or  numerical solutions which would
then give wormholes.

\begin{figure*}[ht]
\centering
\includegraphics[scale=0.6]{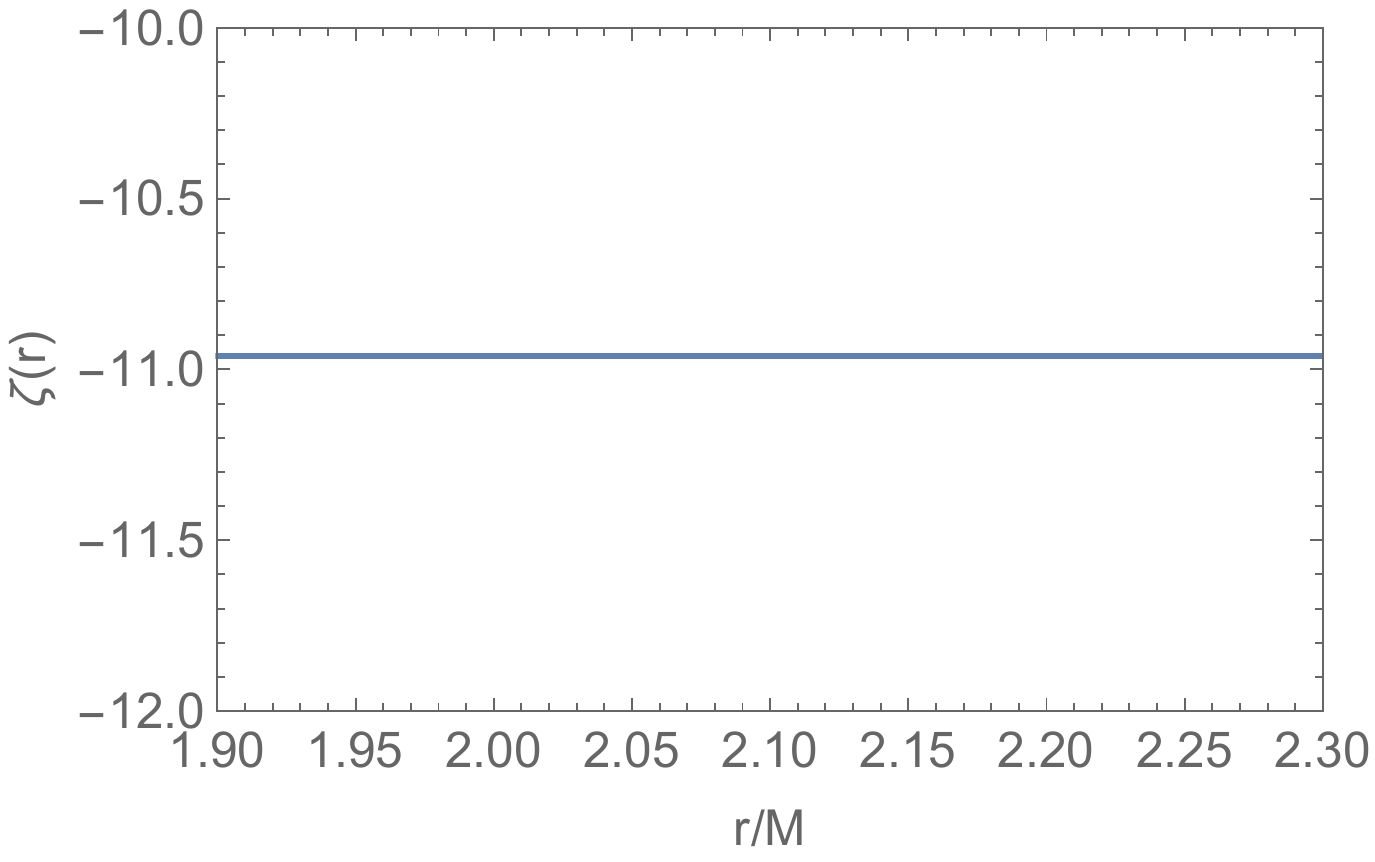}
\includegraphics[scale=0.6]{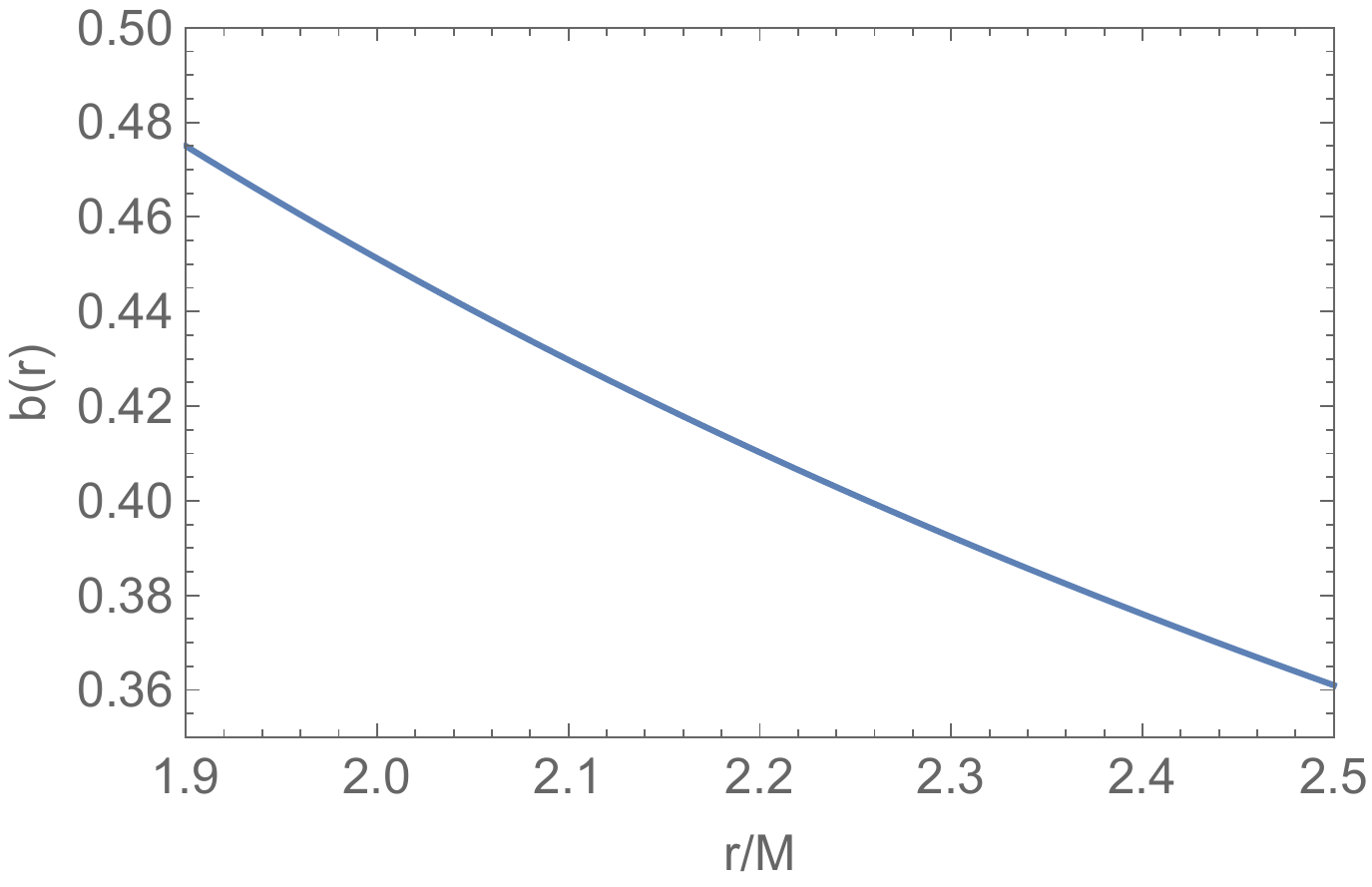}
\includegraphics[scale=0.6]{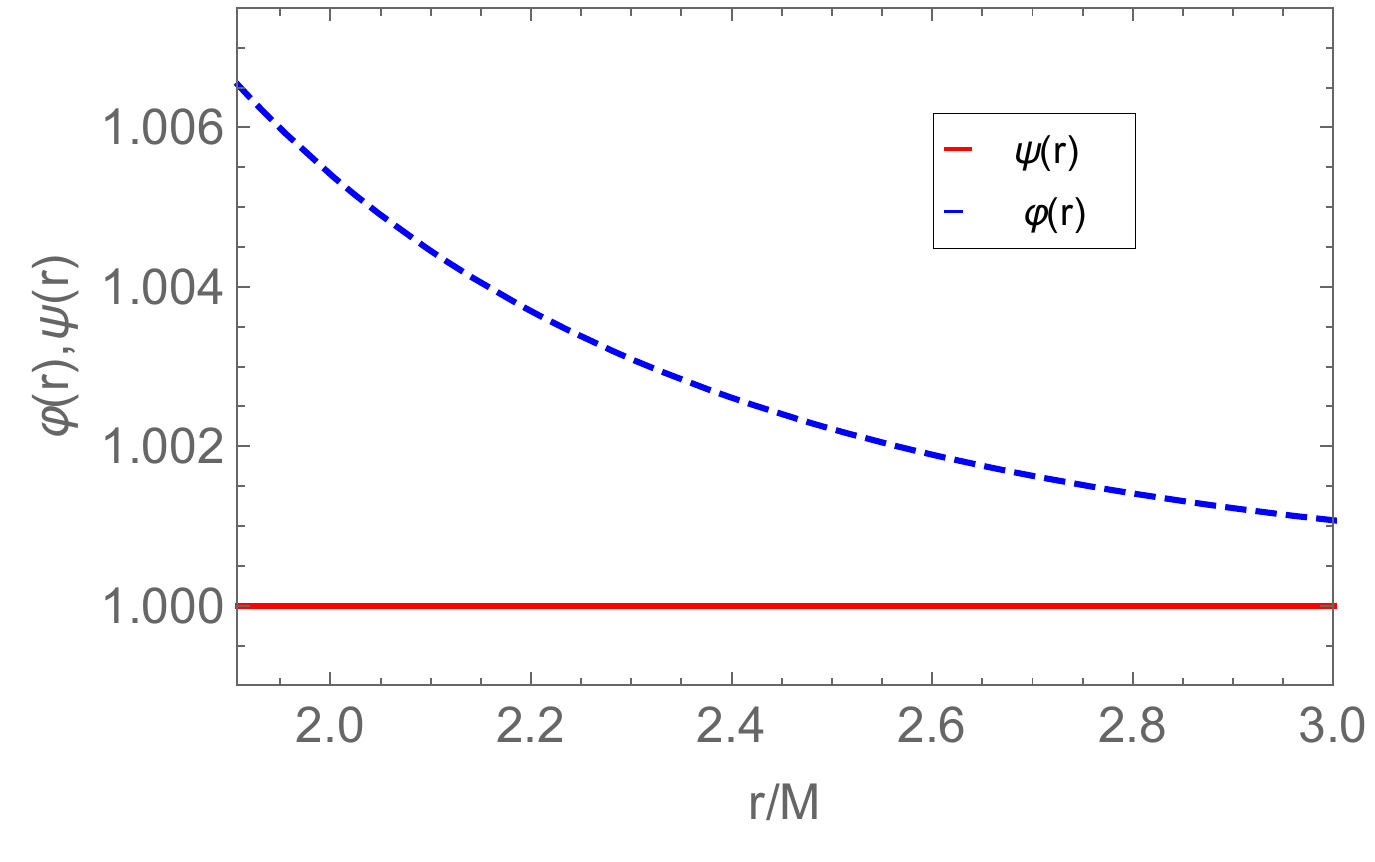}
\includegraphics[scale=0.6]{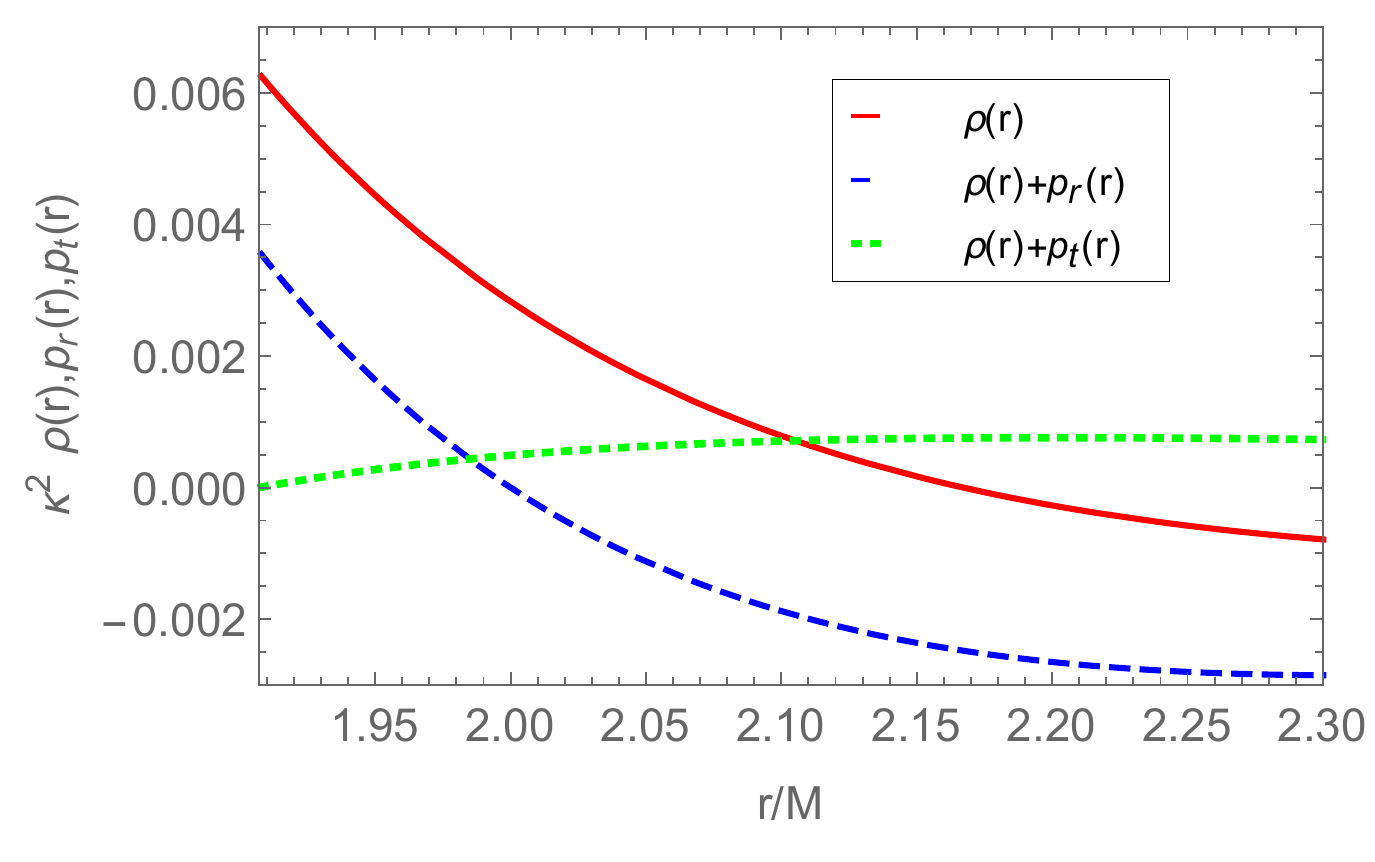}
\caption{Solutions for the metric fields $\zeta(r)$ and $b(r)$, the
scalar fields $\varphi(r)$ and $\psi(r)$, and the matter fields
$\rho$, $p_r$ and $p_t$, where for the latter two it is displayed the
combination of fields that bear directly to the NEC,
i.e., $\rho+p_r$ and $\rho+p_t$.  The values for the parameters are
$\zeta_0=-10.96$, $M=1$, the throat radius is at
$r_0=2\sqrt{10/11}=1.90693$, $\psi_0=1$, $\psi_1=0$, and $V_0=-42$.
Both the scalar fields are finite for all $r$ and converge to the same
constant at infinity, which implies $\varphi-\psi\to 0$ at
infinity. Relative to the NEC, note that $\rho+p_r$
is positive at $r=r_0$ and changes sign at a finite value of $r$,
$r=2$, and
$\rho+p_t$ is positive for every value of $r$.  Note that $\rho$ is
also positive at $r=r_0$ and changes sign at a higher value of $r$,
$r=2.16$,
after which the WEC is violated.  Thus,
the NEC and WEC
are satisfied at the wormhole throat but are
violated somewhere further up.  In order to maintain the validity of
the NEC we have to cut it before it starts to be
violated and match it to an exterior vacuum.  See the text for more
details.}
\label{wormsol1}
\end{figure*}

Here we are interested in finding solutions that do not violate
the matter NEC anywhere, Eqs.~(\ref{rhopr})-(\ref{rhopt}),
in particular at the throat,
$\rho\left(r_0\right)+p_r\left(r_0\right)>0$ and
$\rho\left(r_0\right)+p_t\left(r_0\right)>0$,
and possibly at its neighborhood.
As we will see, a
solution that respects these conditions is
obtained for $\alpha=\gamma=\delta=0$, and $\beta=1$.
For these choices, Eqs.~(\ref{redshift1}) and~(\ref{shape11})
yield 
\be
\label{zeta11}
\zeta\left(r\right)=\zeta_0\,, 
\ee
\be
\label{b11}
b\left(r\right)=\frac{r_0^2}{r}\,, 
\ee
respectively.
Thus, the line element 
Eq.~(\ref{worm}) can be written for the interior
wormhole region containing the throat as
\be
\label{worm2}
ds^2=-e^{\zeta_0}dt^2+\left[1-
\frac{r_0^2}{r^2}\right]^{-1}dr^2+r^2
\left( d\theta^2+\sin^2\theta d\phi^2\right)\,.
\ee
Choosing $\eta=2$ 
Eq.~(\ref{powerpotential})
yields
\be
V\left(\varphi,\psi\right)=V_0\left(\varphi-\psi\right)^2\,.
\label{eqpvvvv}        
\ee
Inserting the metric solutions, Eqs.~\eqref{zeta11}
and~\eqref{b11},
into
Eqs.~\eqref{genkgtrace} and~\eqref{genwormkg1} gives
\be\label{eqphi}
r_0^2+r^4V_0\left(\varphi-\psi\right)=0\,,
\ee
and 
\be\label{eqpsi}
\psi''-\frac{\psi'^2}{2\psi}+\frac{\psi'}{r}
\left[\frac{2-\left(\frac{r_0}{r}\right)^2}{1-
\left(\frac{r_0}{r}\right)^2}\right]=0\,,
\ee
respectively.
Equation \eqref{eqpsi} can be solved analytically to obtain
an expression for the scalar field $\psi\left(r\right)$,
namely,
$\psi\left(r\right)=\left[
\sqrt{\psi_0}+\sqrt{\psi_1} \arctan
\left(\sqrt{\frac{r^2}{r_0^2}-1}\right)\right]^2$,
where $\psi_0$ and $\psi_1$ are integrating constants.
Putting it into Eq.~\eqref{eqphi} we
find the  solution for $\varphi\left(r\right)$. 
Thus, the solutions for 
$\varphi\left(r\right)$ and $\psi\left(r\right)$
are
\be\label{solphi}
\varphi\left(r\right)=\left[\sqrt{\psi_0}+\sqrt{\psi_1 }\arctan
\left(\sqrt{\frac{r^2}{r_0^2}-1}\right)\right]^2-\frac{r_0^2}{r^4V_0}.
\ee
and
\be
\label{solpsi}
\psi\left(r\right)=\left[
\sqrt{\psi_0}+\sqrt{\psi_1} \arctan
\left(\sqrt{\frac{r^2}{r_0^2}-1}\right)\right]^2\,,
\ee
respectively.
We can simplify further and still find nontrivial solutions.
Let us choose  $\psi_1=0$, so that the solution for $\varphi(r)$
in  Eq.~\eqref{solphi} is
\be
\label{phibui}
\varphi(r)=\psi_0-\frac{r_0^2}{r^4V_0}\,, 
\ee
and the solution for $\psi(r)$
in Eq.~\eqref{solpsi} is
\be
\label{psibui}
\psi(r)=\psi_0\,.
\ee
From
Eqs.~\eqref{phibui} and~\eqref{psibui}
we see that both scalar
fields are finite for $r$
at the throat $r_0$,  and for
values of $r$ in its vicinity, and indeed for all
$r$.

So far we have the solution for the fundamental
gravitational fields, displayed in
Eqs.~\eqref{zeta11}
and~\eqref{b11} or in the line element
Eq.~\eqref{worm2},
for the scalar potential,
Eqs.~\eqref{eqpvvvv},
and for the fundamental extra gravitational
fields represented by the scalar fields,
Eqs.~\eqref{phibui} and~\eqref{psibui}.
We can now find the matter fields and study in which
cases they obey the matter NEC.
Inserting these solutions 
into 
Eq.~\eqref{genfieldr},~\eqref{genfieldpr},
and~\eqref{genfieldpt}, we obtain the energy
density, the radial pressure, and the tangential
pressure
as
\be
\rho=\frac{r_0^2}{2r^6V_0\kappa^2}\left(24-31
\frac{r_0^2}{r^2}\right)\,,
\label{indens}
\ee
\be
p_r=\frac{r_0^2}{2r^6V_0\kappa^2}\left(16-13\frac{r_0^2}{r^2}\right)\,,
\ee
\be
p_t=\frac{r_0^2}{2r^6V_0\kappa^2}\left(-32+39\frac{r_0^2}{r^2}\right)\,,
\ee
respectively.

It is more useful to combine $\rho$
with $p_r$ and $p_t$
to obtain expressions directly connected to
the NEC. These combinations give
\be
\label{indenspr}
\rho+p_r=\frac{2r_0^2}{r^6V_0\kappa^2}\left(10-11\frac{r_0^2}{r^2}\right),
\ee
\be
\label{indenspt}
\rho+p_t=\frac{4r_0^2}{r^6V_0\kappa^2}\left(-1+\frac{r_0^2}{r^2}\right).
\ee
At the throat, $r=r_0$, one has
$\rho+p_r=\frac{2}{r_0^4(-V_0)\kappa^2}$.
As we want to preserve the NEC at the
throat we have to set $V_0$ negative,
\be
V_0<0\,.
\label{v0<0}
\ee
In this case 
the NEC 
is valid in the throat's 
neighborhood
up to to the radius $r$
given by 
$r=
\sqrt{\frac{11}{10}}\,r_0
$.
The quantity
$\rho+p_t$ is positive for every value of $r$ and zero at the
throat $r_0$.  Thus the quantity $\rho+p_r$ sets the violation of
the NEC.  We then impose that the interior solution
is valid within the following range of $r$,
\be
r\leq\sqrt{\frac{11}{10}}\,r_0\,.
\label{r1110}
\ee
So, the interior solution must stop at most at
$r=\sqrt{\frac{11}{10}}\,r_0$.
Since $
\rho=\frac{r_0^2}{2r^6V_0\kappa^2}\left(24-31\frac{r_0^2}{r^2}\right)
$, see Eq.~(\ref{indens}),
we have that in this range of $r$ the WEC, i.e.,
$\rho\geq0$, is also
obeyed.

The interior solution
with its metric fields $\zeta(r)$ and $b(r)$,
scalar fields $\varphi(r)$ and $\psi(r)$, and matter
fields $\rho$,
$\rho+p_r$, and 
$\rho+p_t$ are displayed for all $r$ in
Fig.~\ref{wormsol1} for
adjusted values of the free parameters. 
The quantity 
$\rho+p_r$ turns negative at $r/M=2$
which is precisely 
$\sqrt{\frac{11}{10}}\,r_0$
for $r_0/M=2\sqrt{\frac{10}{11}}=
1.90693$. The interior solution
should at most stop there.

\subsubsection{Solutions outside the throat's neighborhood
up to infinity: Schwarzschild spacetimes with a cosmological constant}

To guarantee that the complete solution does obey the NEC
for any value of $r$, we need to
match at some $r_\Sigma$, say, less than $\sqrt{\frac{11}{10}}\,r_0$ the
interior solution just found to an external vacuum spherically symmetric solution.
To do so, one has to derive a vacuum
and asymptotically flat exterior solution, and then use the junction
conditions of this theory to perform the matching with some thin shell
at the boundary between the interior and exterior.  Having performed
this, we still have to ensure that the matter of this thin shell obeys
the matter NEC, in order that the complete
spacetime obeys theis condition.

We now find an exterior vacuum solution so that we can use
the junction conditions to match the
interior wormhole solution to the exterior vacuum solution. To do so,
we put the stress-energy tensor
to zero, $T_{ab}=0$,
as we want a vacuum solution.
In addition, we chose the scalar fields
to be constant in this exterior solution,
i.e.,
$\varphi\left(r\right)=\varphi_e$
and $\psi\left(r\right)=\psi_e$
with
$\varphi_e$ and
$\psi_e$ constants, $\varphi_e\neq\psi_e$, and where the
subscript $e$ stands for exterior.
 For continuity we choose the potential
to be
$V=V_0\left(\varphi_e-\psi_e\right)^2$.
Note that from Eq.~\eqref{potential} it can be seen that choosing a
particular form of the potential $V$ we are also setting a particular
solution for the function $f$. This means that both the interior and
the exterior spacetimes must be solutions of the field equations with
the same form of the potential $V$, because otherwise they would not
be solutions of the same form of the function $f\left(R,\mathcal
R\right)$. These choices imply
that the field equation Eq.~\eqref{genein2} can be
written as
\be
G_{ab}+\frac{V_0}{2}\left(\varphi_e-\psi_e\right)g_{ab}=0\,.
\ee
Considering that in the second term the multiplicative factors of the
metric $g_{ab}$ are constant, we see that the field
equation is of the same form as the
Einstein field equations in vacuum with a cosmological constant
given by
$\frac{V_0}{2}\left(\varphi_e-\psi_e\right)$.
Thus,
the Schwarzschild so\-lution
with a cosmological constant of general relativity
is a vacuum solution of the
generalized hybrid
theory we are studying.
This class of solutions is also known as
the Kottler solution, as well as  Schwarzschild-dS
solution if the constant cosmological term is positive
and Schwarzschild--AdS
solution if the constant cosmological term is negative.
The metric fields $\zeta(r)$ and $b(r)$ for the
exterior region outside some radius $r_\Sigma$
are then
\be
\label{zeta1e}
e^{\zeta(r)}=\left(
1-\frac{2M}{r}-
\frac{V_0\left(\varphi_e-\psi_e\right)r^2}{6}
\right)
e^{\zeta_e}\,, 
\ee
\be
\label{b1e}
b\left(r\right)=2M+
\frac{V_0\left(\varphi_e-\psi_e\right) r^3}{6}\,, 
\ee
respectively.
where $M$ is a constant
of integration and represents the mass, and the factor
$e^{\zeta_e}$ is a
useful constant.
The line element  of the generalized field equations given
by Eq.~\eqref{genein2} in the case where the scalar fields
are constant
is then
\beq\label{kottler}
ds^2=&&-\left({1-\frac{2M}{r}-
\frac{V_0\left(\varphi_e-\psi_e\right) r^2}{6}}
\right)e^{\zeta_e}
dt^2+
\\
&&+\left(1-\frac{2M}{r}-\frac{V_0\left(\varphi_e-\psi_e\right)
r^2}{6}\right)^{-1}dr^2+r^2d\Omega^2\,,\nonumber\\
&&r>r_\Sigma\nonumber\,,
\eeq
The sign of the
term $\frac{V_0}{2}\left(\varphi_e-\psi_e\right)$
will
determine whether the  solution is Schwarzschild-dS
or Schwarzschild-AdS.
This sign will be determined by
the matching surface and the imposition that
the NEC holds everywhere.
Wormholes
in dS and AdS spacetimes in general relativity were
treated in \cite{lemoslobooliveira}.

To complete, the potential, scalar fields, and matter fields
have then for the outer solution
the following expressions,
\be
V=V_0\left(\varphi_e-\psi_e\right)^2\,,
\ee
\be
\varphi\left(r\right)=\varphi_e\,,
\ee
\be
\psi\left(r\right)=\psi_e\,,
\ee
with
$\varphi_e$ and
$\psi_e$ constants, and
\be
\rho(r)=0\,,\quad p_r(r)=0\,,\quad p_t(r)=0\,,
\ee
respectively.


\subsubsection{The shell at the junction: the surface density
and pressure of the thin shell}

To match the interior to the exterior solution we need the junction
conditions for the generalized hybrid metric-Palatini gravity.  The
deduction of the junction
equations has been performed in \cite{rosalemos1}.  There are seven
junction conditions in this theory that must be respected in order to
perform the matching between the interior and exterior solutions, two
of them that imply the existence of a thin shell of matter at the
junction radius $r_\Sigma$. These conditions are
\beq
&&\left[h_{\alpha\beta}\right]=0 \label{hab2}\,,\\
&&\left[\varphi\right]=0 \label{phicont}\,,\\
&&\left[\psi\right]=0 \label{psicont}\,,\\
&&n^a\left[\partial_a\psi\right]=0 \label{psider}\,, \\
&&n^a\left[\partial_a\varphi\right]=\frac{\kappa^2}{3}S\,,
\label{junction}\\
&&S_\alpha^\beta-\frac{1}{3}\delta_\alpha^\beta
S=-\frac{\left(\varphi_\Sigma-\psi_\Sigma\right)}{\kappa^2}
\left[K_\alpha^\beta\right]\label{last}\,,\\
&&\left[K\right]=0\,,\label{K=0}
\eeq
where $h_{\alpha\beta}$ is the induced metric at the junction
hypersurface $\Sigma$, with  Greek indexes
standing for $0$ and $2,3$, the
brackets $\left[X\right]$ denote the jump of any quantity $X$ across
$\Sigma$,
$n^a$ is the unit normal vector to $\Sigma$,
$K=K_\alpha^\alpha$ is the trace of the extrinsic curvature
$K_{\alpha\beta}$ of the surface $\Sigma$
(see, e.g., \cite{Visser:1995cc}), 
$S=S_\alpha^\alpha$ is the trace of the stress-energy tensor
$S_{\alpha\beta}$ of the thin shell, $\delta_\alpha^\beta$ is the
Kronecker delta, and the subscripts $\Sigma$ indicate the value
computed at the hypersurface.

Now, the matter NEC
given in
Eqs.~\eqref{rhopr}-\eqref{rhopt}
should also apply to the shell, since we want
this condition to be valid throughout the
whole spacetime. 
When applied for the particular case
of the thin shell the matter NEC is
\be
\label{shellnec}
\sigma+p\geq0\,.
\ee
We have now to construct the shell out of the conditions 
Eqs.~\eqref{hab2}-\eqref{shellnec}.

The condition Eq.~\eqref{hab2} gives, on using 
Eqs.~\eqref{zeta11} and \eqref{zeta1e},
that we can choose
\be
\label{zeta1esigma}
e^{\zeta_e}=
\frac{e^{\zeta_0}}
{1-\frac{2M}{r_\Sigma}-
\frac{V_0\left(\varphi_e-\psi_e\right) r_\Sigma^2}{6}}.
\ee
This factor $e^{\zeta_e}$ has been chosen
so that the
time coordinate $t$
for the interior, Eq.~\eqref{worm2}, is the same as the coordinate $t$
for the exterior, Eq.~\eqref{kottler}.
The angular part of the metrics Eq.~\eqref{worm2}
and Eq.~\eqref{kottler}
are continuous. So the line element at the
surface $\Sigma$ and outside it is
\beq\label{kottler2}
ds^2=&&-\left(\frac{1-\frac{2M}{r}-
\frac{V_0\left(\varphi_e-\psi_e\right) r^2}{6}}
{1-\frac{2M}{r_\Sigma}-
\frac{V_0\left(\varphi_e-\psi_e\right)
r_\Sigma^2}{6}}\right)e^{\zeta_0}dt^2+\\
&&+\left(1-\frac{2M}{r}-\frac{V_0\left(\varphi_e-\psi_e\right)
r^2}{6}\right)^{-1}dr^2+r^2d\Omega^2\,,\nonumber\\
&&r\geq r_\Sigma\,, 
\eeq
where we are using that the $\varphi$ and $\psi$ are
continous at the matching, see below.

The condition Eq.~\eqref{phicont} means that
\be
\label{phicontsigma}
\varphi_e=\varphi_\Sigma=\psi_0-\frac{r_0^2}{r_\Sigma^4V_0}\,,
\ee
where we have used Eq.~\eqref{phibui}.
Whenever it occurs we keep the notation $\varphi_e$.

The condition Eq.~\eqref{psicont} means that
\be
\label{psicontsigma}
\psi_e=\psi_\Sigma=\psi_0\,.
\ee
where we have used Eq.~\eqref{psibui}.
Whenever it occurs we keep the notation $\psi_e$.

The condition Eq.~\eqref{psider} is empty, since
$\psi$ is constant everywhere.

The condition Eq.~\eqref{junction} gives
$\partial_r\varphi|_\Sigma
=-\frac{\kappa^2}{3}S$, 
$n^a\partial_a\varphi_e=0$, and so,
$
n^a\left[\partial_a\varphi\right]=-\partial_r\varphi|_\Sigma
$. All the contribution comes from the interior.
Thus, Eq.~\eqref{junction} yields finally 
\be
\frac{4r_0^2}{r_\Sigma^5V_0}
=-\frac{\kappa^2}{3}S\,.
\label{a11}
\ee

The condition Eq.~\eqref{last} can be used to detemine 
the surface energy density $\sigma$ and the surface transverse pressure
$p$.
Note that $\sigma$ is
given by $S_0^0=-\sigma$ and $p$  is
given by $S_1^1=S_2^2=p$.
Then, from Eq.~\eqref{last} we have 
$\sigma=\frac{1}{\kappa^2}\left(\varphi_\Sigma-
\psi_\Sigma\right)\left[K_0^0\right]-\frac{1}{3}S$,
and 
$p=-\frac{1}{\kappa^2}\left(\varphi_\Sigma-
\psi_\Sigma\right)\left[K_0^0\right]+\frac{1}{3}S$.
For the
interior solution given in  Eq.~\eqref{worm2}
and the exterior solution
given in  Eq.~\eqref{kottler}
we can compute $\left[K_0^0\right]$, 
$
\left[K_0^0\right]=\frac{r_0\zeta_0}{2r_\Sigma^2}\sqrt{1-
\frac{r_0^2}{r_\Sigma^2}}-\frac{r_\Sigma^3V_0
\left(\varphi_e-\psi_e\right)-6M}{6r_\Sigma^2
\sqrt{1-\frac{2M}{r_\Sigma}-\frac{r_\Sigma^2}{6}V_0
\left(\varphi_e-\psi_e\right)}}
$, which upon using 
 Eq.~\eqref{phicontsigma},
\eqref{psicontsigma}, and 
\eqref{a11} gives 
\be\label{extr2}
\left[K_0^0\right]=\frac{r_0\zeta_0}{2r_\Sigma^2}\sqrt{1-
\frac{r_0^2}{r_\Sigma^2}}+\frac{\frac{r_0^2}{r_\Sigma}-6M}{6r_\Sigma^2
\sqrt{1-\frac{2M}{r_\Sigma}+\frac{r_0^2}{6r_\Sigma^2}}}\,. 
\ee
Then, $\sigma$ and $p$ are  given by
\be
\label{dens2}
\sigma=\frac{4r_0^2}{\kappa^2V_0r_\Sigma^5}\left(1-
\frac14\, r_\Sigma
\left[K_0^0\right]\right)\,,
\ee
\be
\label{pres2}
p=-\frac{4r_0^2}{\kappa^2V_0r_\Sigma^5}\left(1+
\frac18\, r_\Sigma
\left[K_0^0\right]\right)\,,
\ee
respectively.

The condition Eq.~\eqref{K=0} means that the matching
has to be performed at the radius $r_\Sigma$ such
that the jump in the trace of the extrinsic curvature
is zero, i.e.,
$
\left[K\right]=\frac{1}{r_\Sigma}\left[\frac{r_0\zeta_0}{2r_\Sigma}
\sqrt{1-\left(\frac{r_0}{r_\Sigma}\right)^2}-
\sqrt{1-\left(\frac{r_0}{r_\Sigma}\right)^4}
-\sqrt{1-\left(\frac{r_0}{r_\Sigma}\right)^5}+\right. $

\noindent $
\left.\frac{2-\frac{3M}{r_\Sigma}-\frac{1}{2}r_\Sigma^2V_0
\left(\varphi_e-\psi_e\right)}{\sqrt{1-
\frac{2M}{r_\Sigma}-\frac{1}{6}r_\Sigma^2V_0\left(\varphi_e-
\psi_e\right)}}\right]=0
$.
Upon using 
Eqs.~\eqref{phicontsigma},
\eqref{psicontsigma}, and 
\eqref{a11} it gives
\beq
\label{trace2}
&&\left[K\right]=\frac{1}{r_\Sigma}\left[\frac{r_0\zeta_0}{2r_\Sigma}\sqrt{1-\left(\frac{r_0}{r_\Sigma}\right)^2}-\sqrt{1-\left(\frac{r_0}{r_\Sigma}\right)^4}\right. \nonumber\\
&&\left.-\sqrt{1-\left(\frac{r_0}{r_\Sigma}\right)^5}+\frac{2-\frac{3M}{r_\Sigma}+\frac{r_0^2}{2r_\Sigma^2}}{\sqrt{1-\frac{2M}{r_\Sigma}+\frac{r_0^2}{6r_\Sigma^2}}}\right]=0\,.
\eeq

The condition Eq.~\eqref{shellnec}
is the matter NEC
that should be valid on the shell. In this way
we can keep the validity of the matter NEC
throughout the
whole spacetime. 
The relevant quantity is
$\sigma+p$. Using Eqs.~(\ref{dens2}) and ~(\ref{pres2})
we find
\be
\label{ns1}
\sigma+p=
-\frac{3r_0^2}{2\kappa^2V_0r_\Sigma^4}\left[K_0^0\right]
\,.
\ee

Finding a combination of parameters that fulfills all the junction
conditions, 
plus the matter NEC everywhere,
is a problem that
requires some fine-tuning. We now turn to this.


\subsection{The full wormhole solution obeying the matter null
energy condition everywhere}

\begin{figure*}[ht]
\centering
\includegraphics[scale=0.6]{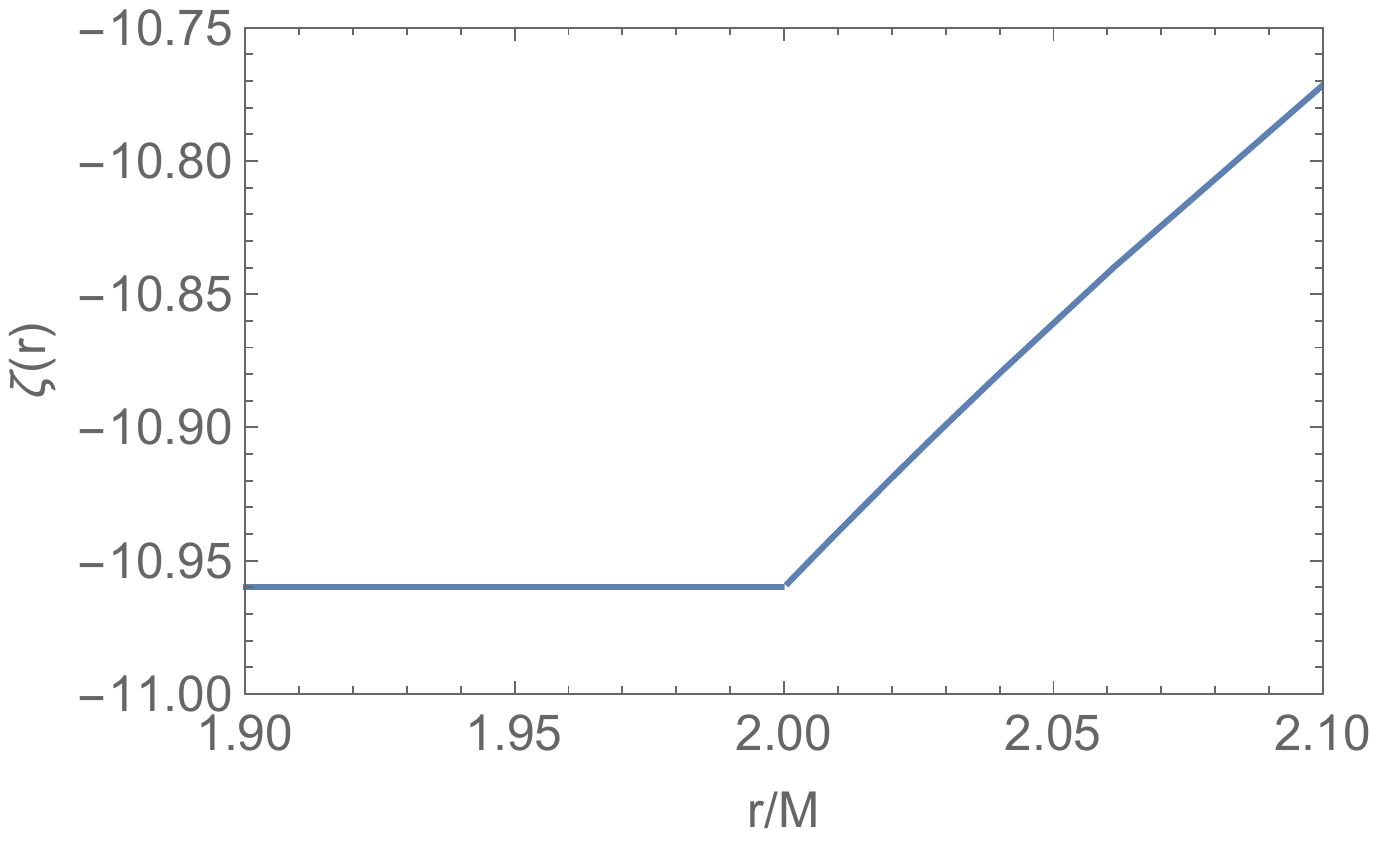}
\includegraphics[scale=0.6]{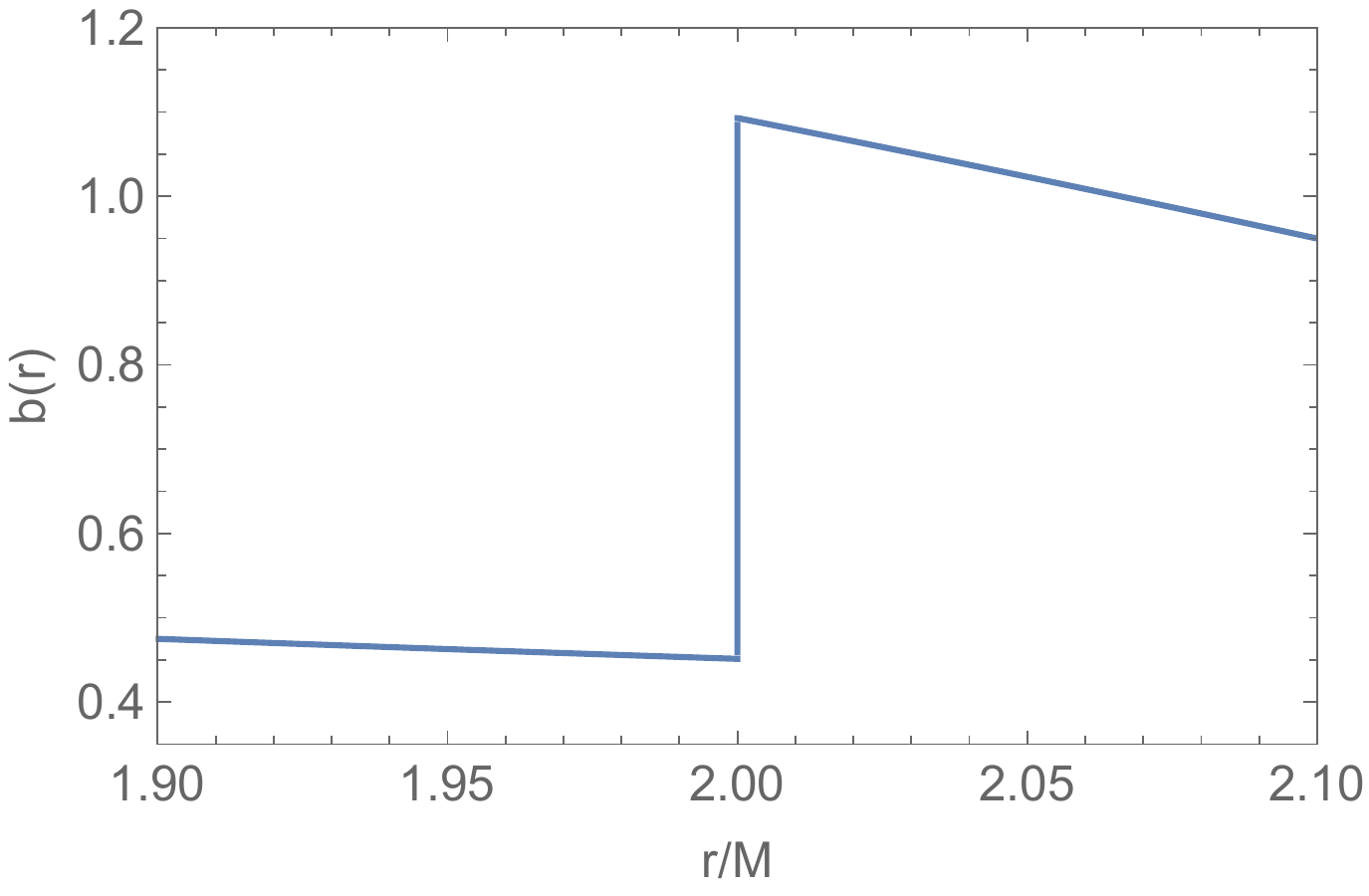}
\includegraphics[scale=0.6]{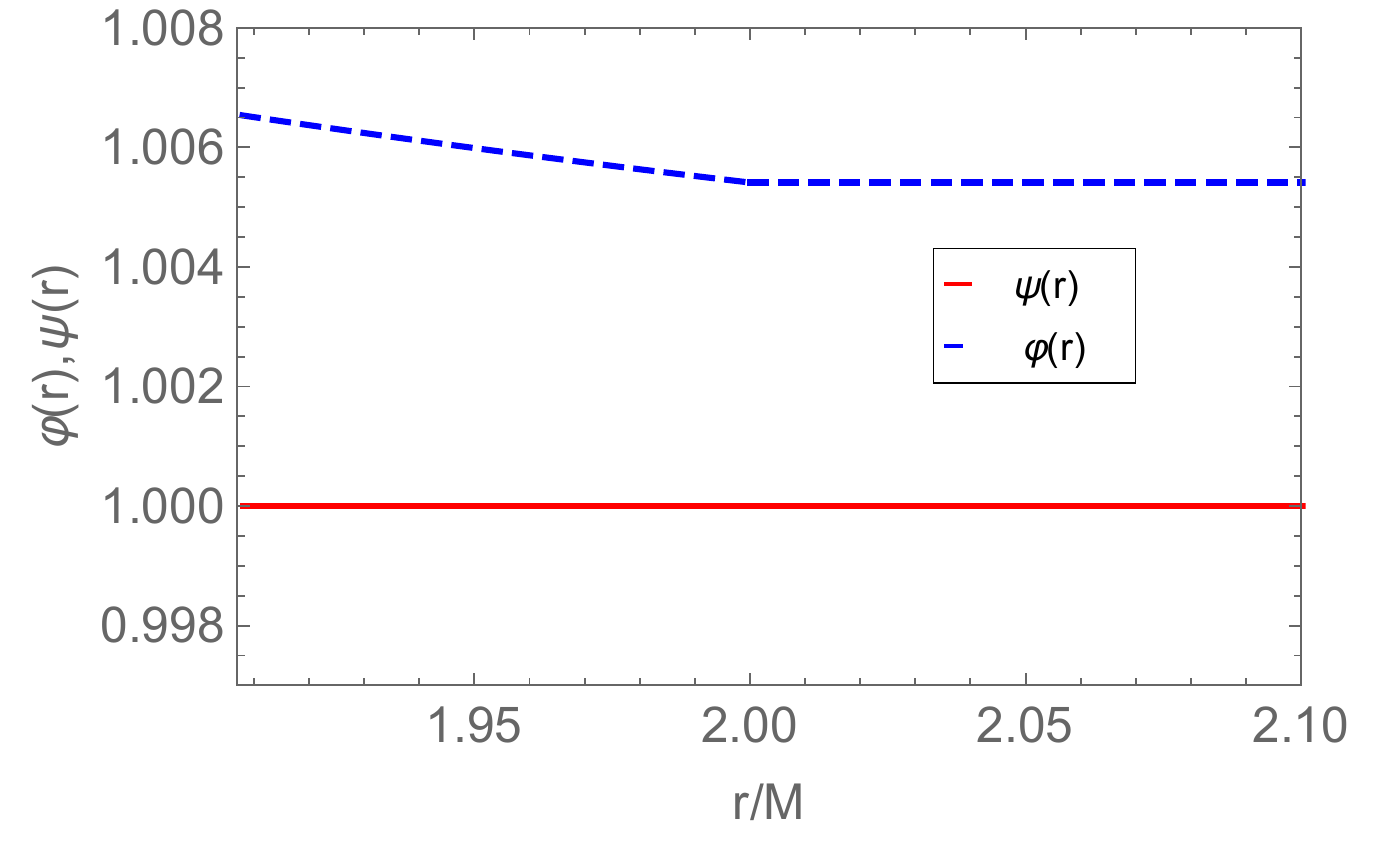}
\includegraphics[scale=0.6]{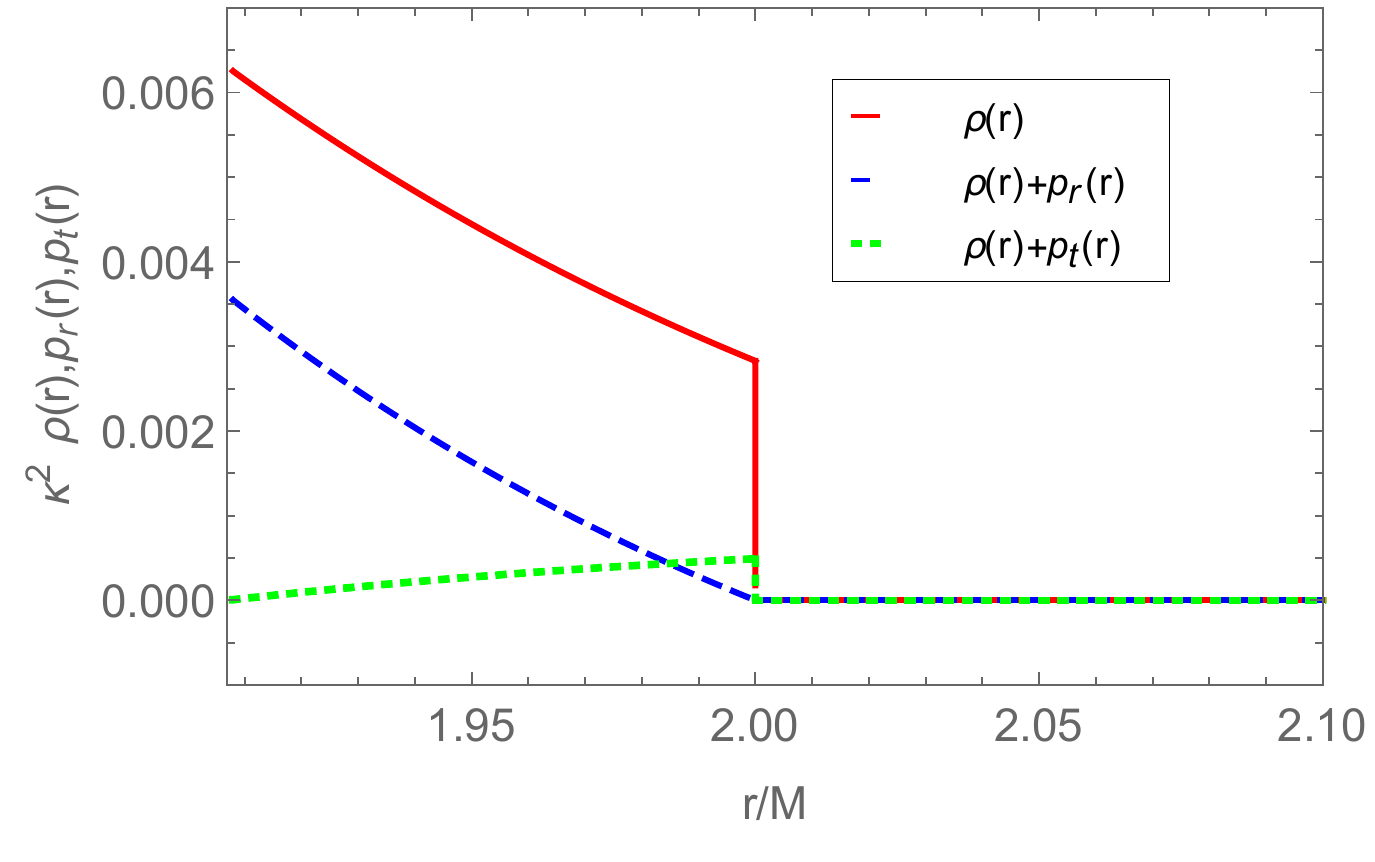}
\caption{
Full wormhole solution from the throat to infinity with the
NEC and WEC
being obeyed everywhere.  The solutions are for
the metric fields $\zeta(r)$ and $b(r)$, the scalar fields
$\varphi(r)$ and $\psi(r)$, and the matter fields $\rho$, $\rho+p_r$
and $\rho+p_t$.  The values of the free parameters are
$\zeta_0=-10.96$, $M=1$, the throat is at $r_0=2\sqrt{10/11}=1.907$,
$\psi_0=1$, $\psi_1=0$, $V_0=-42$, and
the matching surface is at
$r_\Sigma=2$.  The metric
fields $\zeta(r)$ and $b(r)$ are asymptotically AdS, and a
thin shell of matter is perceptibly present at the matching surface
$r_\Sigma=2$ (more properly
at $r_\Sigma=2M$ in our solution, and here we put $M=1$), thus outside
the gravitational radius of the solution.  Both the scalar fields are
finite for all $r$ and converge to the same constant at infinity,
which implies $\varphi-\psi\to 0$ at infinity.
Relative to the
NEC, note
that 
$\rho+p_r$ is
positive at $r=r_0$
and up to $r=2$ and for $r>2$ is zero.
The quantity
$\rho+p_t$ is zero at $r=r_0$,
positive for every other value of $r<2$,
and for $r>2$ is zero. At the shell $r=2$
one has $\kappa^2(\sigma+p)=\frac{0.02}{(-V_0)}
=0.0005$ (not displayed). 
Thus the NEC is satisfied everywhere.
See the text for more details.
}
\label{wormsol2}
\end{figure*}

First, the solution scales with the exterior mass $M$ appearing in
Eq.~\eqref{kottler} and so all quantities can be normalized to it.

Second, the parameter $V_0$, as 
we already argued, must be negative, $V_0<0$,
so that the NEC is verified at the throat
and its vicinity, see Eq.~\eqref{indenspr}.
In this case 
the WEC is also verified at the throat
and its vicinity.

Third, the junction condition $\left[K\right]=0$, see
Eq.~\eqref{trace2}, can be achieved only for certain values of the
radius $r\equiv r_\Sigma$.  We certainly need to guarantee that the
radius $r_\Sigma$ at which $\left[K\right]=0$ happens is inside the
region $r<\sqrt{11/10}\,r_0$, see Eq.~\eqref{r1110}, so that the
interior solution does not violate the NEC. Manipulation of the parameters of the interior solution
shows that these conditions can be achieved by changing the parameter
$r_0$.

Fourth, once we have set values for $r_0$ and $r_\Sigma$,
Eq.~\eqref{trace2} automatically sets the value of the parameter
$\zeta_0$ and hence, Eq.~\eqref{extr2} sets the value of
$\left[K_0^0\right]$. Then, by Eqs.~\eqref{dens2} and \eqref{pres2},
we see that choosing a value of the parameter $V_0$ determines the
values of $\sigma$ and $p$. However, since we already argued that
$V_0$ must be negative in order for the NEC
to be verified at the throat, see Eq.~\eqref{indenspr}, then
the signs of $\sigma$ and $p$ are determined even without specifying
an exact value for $V_0$.
Moreover, 
since $V_0<0$, the NEC at the shell, Eq.~(\ref{ns1}),
is satisfied if $\left[K_0^0\right]\geq0$.

Fifth, 
having all the parameters determined, one has in general to verify
that $r_\Sigma$ is greater that the gravitational radius $r_g$ of the
solution. If this were not the case then there would be a horizon and
the solution would be invalid. In our solution, one has
$r_g=2M\left[1+2V_0(\varphi_e-\psi_e)M^2/3\right]$, see
Eq.~\eqref{kottler}. However, using Eqs.~\eqref{phicontsigma} and
\eqref{psicontsigma}, one can write the gravitational radius as
$r_g=2M\left(1-\frac{2M^2r_0^2}{3r_\Sigma^4}\right)$, which is smaller
than $r_\Sigma$ for any value of $r_\Sigma$ between $r_0$ and
$\sqrt{\frac{11}{10}}r_0$. This implies that in the range of solutions
that we are interested in this step is automatically satisfied.

We are now in a position to specify a set of parameters for which
the matter
NEC is
obeyed everywhere, i.e., in the interior,  on
the shell, and in the exterior, the latter being a trivial case since
there is no matter in this region. We give an example and consider a
wormhole with mass $M$.
Note that 
the conclusions are the same for any combination of $r_0$ and $r_\Sigma$
within the allowed region for these two parameters.
A concrete convenient example is to 
have a  wormhole interior region
for which the radius $r_0$ of the throat has the value
$r_0=\sqrt{\frac{10}{11}}2M$.
Then, from Eq.~\eqref{r1110}
we can put
$r_\Sigma=2M$ and have the matter NEC
in the interior being
obeyed.
With these values of 
$r_0$ and 
$r_\Sigma$,
Eq.~\eqref{trace2} sets the value of $\zeta_0$ to be
$\zeta_0=-10.96$.  From Eq.~\eqref{extr2} we get
$\left[K_0^0\right]= 0.049$, and using Eqs.~\eqref{dens2} and
\eqref{pres2} we compute the values of the stress-energy tensor at the
shell, which become $\kappa^2\sigma= \frac{0.44}{V_0}$ and
$\kappa^2 p= -\frac{0.46}{V_0}$, respectively, and hence
$\kappa^2\left(\sigma+p\right)= -\frac{0.02}{V_0}$.  Since
$V_0<0$ we see that $\sigma+p>0$ and thus the NEC is
satisfied at the shell.
Since $\kappa^2\sigma= \frac{0.44}{V_0}$ and 
$V_0<0$ we have $\sigma<0$.
So the  WEC
does not hold on the shell but holds everywhere else.
The NEC holds everywhere.

This completes our full wormhole solution.
Fig.~\ref{wormsol2} displays a solution with $M=1$,
$r_0=\sqrt{\frac{10}{11}}2$, $r_\Sigma=2$, $\zeta_0=-10.96$,
$V_0=-42$.
This wormhole solution obeys the matter NEC
everywhere in conformity with our aim.

Other wormhole solutions in this
generalized hybrid theory with other choices of parameters, within the
ranges stated above, that obey the matter NEC
everywhere can be found along the lines we have presented. 

\section{Conclusion}\label{conclusion}

In this work, we found traversable asymptotically AdS wormhole
solutions that obey the NEC everywhere in the generalized hybrid
metric-Palatini gravity theory, so there is no need for exotic matter.
The generalized hybrid theory consists in a gravitational action given
by $f\left(R,\cal{R}\right)$, where $R$ is the metric Ricci scalar,
and $\mathcal{R}\equiv\mathcal{R}^{ab}g_{ab}$ is the Palatini
curvature defined in terms of an independent connection, to which a
matter action is added.  The gravitational action can be given in the
scalar-tensor representation where the metric Ricci scalar $R$ is
still present but now coupled to two scalar fields $\varphi$ and
$\psi$. The equations of motion in this representation were obtained.

The interior wormhole solution obtained verifies the NEC near and at
the throat of the wormhole.  The matching of the interior solution to
an exterior vacuum solution yields a thin shell respecting also the
NEC.  Finding a combination of parameters that allow for the interior
and the exterior solution to be matched without violating the NEC in
the interior solution and at the thin shell is a problem that requires
fine-tuning, but can be acomplished. We presented a specific
combination of parameters with which it is possible to build the full
wormhole solution and found that it is asymptotically AdS.  Most of
the work that has been done in hybrid metric-Palatini gravity theories
has been aimed to find  solutions where the NEC is satisfied solely at
the throat of the wormhole. Within these theories our solution is the
first where the NEC is verified for the entire spacetime.

There are two main interesting conclusions. On one hand, the existence
of these solutions is in agreement with the understanding that traversable
wormholes supported by extra fundamental gravitational fields, here in
the guise of scalar fields, can occur without the need of exotic
matter. On the other hand, the rather contrived construction necessary
to build a spacetime complete wormhole solution may indicate that
there are not many such solutions around in this class of theories.

\centerline{}
\vskip 1cm
\begin{acknowledgments}
JLR acknowledges financial support of FCT-IDPASC through grant
no. PD/BD/114072/2015.
JPSL thanks Funda\c c\~ao para a Ci\^encia e Tecnologia (FCT), Portugal,
for financial support through Grant~No.~UID/FIS/00099/2013
and Grant No.~SFRH/BSAB/128455/2017, and Coordena\c c\~ao de
Aperfei\c coamento do Pessoal de N\'\i vel Superior (CAPES),
Brazil, for support within the Programa CSF-PVE, Grant
No.~88887.068694/2014-00. 
FSNL acknowledges
financial support of the FCT through an Investigador FCT Research
contract, with reference IF/00859/2012, and the grant
PEst-OE/FIS/UI2751/2014.
\end{acknowledgments}


\end{document}